\documentclass[epj]{svjour}

\usepackage{graphicx}
\usepackage{fancyhdr}
\def\makeheadbox{{
 }}
\def\mytitle{My title} 
\def\myauthors{My name}  
\def\mytype{My type of session}
\def\mysession{My session}

\def\mytitle{The Lightest Higgs Boson and Relic Neutralino in the MSSM with CP Violation} 
\def\myauthors{Stefano Scopel}    
\def\mytype{Contributed Talk}    
\def\mysession{Cosmology and Astrophysics}

\newcommand{\gsim}{\lower.7ex\hbox{$\;\stackrel{\textstyle>}{\sim}\;$}}
\newcommand{\lsim}{\lower.7ex\hbox{$\;\stackrel{\textstyle<}{\sim}\;$}}

\begin{document}
\title{Lightest Higgs Boson and Relic Neutralino in the MSSM with CP Violation}
\author{Stefano Scopel\inst{1}
\thanks{\emph{Email: scopel@kias.re.kr}}%
 \and
 Jae Sik Lee\inst{2}
}                     

%
\institute{Korea Institute for Advanced Study, Seoul
130-722, Korea
\and Center of Theoretical Physics, School of
Physics, Seoul National University, Seoul, 151--747, Korea}
%
\date{}
\abstract{ 
In a minimal supersymmetric extension of the standard model
(MSSM) with explicit CP violation a mass of the lightest Higgs boson
$H_1$ in the range $7~{\rm GeV}\lsim
M_{H_1}\lsim 10$ GeV is experimentally allowed by present accelerator
limits. In the same scenario a lightest neutralino as light as 2.9 GeV
can be a viable dark matter candidate, provided that a  departure from
the usual GUT relation among gaugino masses is assumed.
\PACS{
      {12.60.Jv,}{Supersymmetric models}   \and
      {14.80.Cp}{Non-standard-model Higgs bosons} \and
      {95.35.+d}{Dark matter}
     } 
} 
\maketitle
\section{Introduction}
\label{intro}

In SUSY models, CP violating phases in the soft terms can considerably
enrich the phenomenology without violating existing constraints.  In
particular, in such a scenario it is well known that the mass of the
lightest Higgs boson $H_1$ can be much lighter compared to the CP
conserving case.  We wish here to take a fresh look at how light
$m_{H_1}$ can be, and discuss the cosmological lower bound to the mass
of the relic neutralino $\chi$, when standard assumptions are made for
the origin and evolution of its relic density. For more details of
this analysis, see \cite{scopel_lee}.

\section{MSSM with explicit CP violation: the CPX scenario}

In the presence of sizable CP phases in the relevant soft SUSY breaking
terms, a significant mixing between the scalar and pseudo--scalar
neutral Higgs bosons can be generated in the Higgs potential at one loop
~\cite{CPsuperH_ELP}.  As a consequence of this,
the three neutral MSSM Higgs mass eigenstates, labeled in order of
increasing mass as $M_{H_1}\le M_{H_2} \le M_{H_3}$, have no longer
definite CP parities, but become mixtures of CP-even and CP-odd
states. 
Due to the large Yukawa couplings, the CP-violating mixing among the
neutral Higgs bosons is dominated by the contribution of
third-generation squarks and is proportional to the combination:
\begin{equation}
\frac{3}{16\pi^2}\frac{\Im{\rm
m}(A_f\,\mu)}{m_{\tilde{f}_2}^2-m_{\tilde{f}_1}^2}\label{eq:ratio},
\end{equation}
\noindent with $f=t,b$. Here $\mu$ is the Higgs--mixing parameter in
the superpotential and $A_f$ denotes the trilinear soft coupling.  

In presence of CP violation, the mixing among neutral Higgs bosons is
described by a 3$\times$3 real orthogonal matrix $O$:
\begin{equation}
(\phi_1\,,\phi_2\,,a)^{T}=O\,(H_1\,,H_2\,,H_3)^T,
\label{eq:omix}
\end{equation}
\noindent 
where the elements $O_{\phi_1 i}$ and $O_{\phi_2 i}$ are 
the CP-even components of the $i$-th Higgs
boson, while $O_{a i}$ is the corresponding CP-odd component.

The Higgs-boson couplings to the SM and SUSY particles can be modified
significantly due to the CP violating mixing.  Among them, one of the
most important ones is the Higgs-boson coupling to a pair of vector
bosons, $g_{H_iVV}$, which is responsible for the production of Higgs
bosons at $e^+e^-$ colliders:
\begin{equation}
{\cal L}_{HVV}=gM_W\left(W_\mu^+W^{-\mu}+\frac{1}{2c_W^2}Z_\mu Z^\mu\right)
\sum_{i=1}^3 g_{H_iVV}H_i\,,
\label{eq:hvvinterx}
\end{equation}
where
\begin{equation}
g_{H_iVV}=c_\beta \, O_{\phi_1i}+s_\beta \, O_{\phi_2 i}\,,
\end{equation}
\noindent when normalized to the SM value. Here we have used the following abbreviations:
$s_\beta\equiv\sin\beta$, $c_\beta\equiv\cos\beta$.
$t_\beta=\tan\beta$, etc.
We note that the two vector bosons $W$ and $Z$ couple only to the
CP-even components $O_{\phi_{1,2} i}$ of the $i$-th Higgs mass
eigenstate, and the relevant couplings may be strongly
suppressed when the $i$-th Higgs boson is mostly CP-odd, $O^2_{ai}\sim
1 \gg O_{\phi_1i}^2\,,O_{\phi_2i}^2$.

The so called CPX scenario is defined as a showcase benchmark
point for studying CP-violating Higgs-mixing phenomena ~\cite{CPX}. Its
parameters are all defined at the electro--weak scale, and are chosen
in order to enhance the combination in Eq.(\ref{eq:ratio}).
In this scenario,
SUSY soft parameters are fixed as follows:
\begin{eqnarray}
&& 
M_{\tilde{Q}_3} = M_{\tilde{U}_3}= M_{\tilde{D}_3} =\nonumber\\
&& M_{\tilde{L}_3}
=M_{\tilde{E}_3} = M_{\rm SUSY};|\mu|=4 M_{\rm SUSY},
\nonumber \\
&&  |A_{t,b,\tau}|=2 M_{\rm SUSY}, 
|M_3|=1 ~~{\rm TeV}\,,
\label{eq:mssm-intro-CPXdef}
\end{eqnarray}
\noindent where, with a usual notation, $Q$, $L$, $U$, $D$ and $E$
indicate chiral supermultiplets corresponding to left-- and
right--handed quarks and leptons. In this scenario $\tan\beta$,
$M_{H^\pm}$, and $M_{\rm SUSY}$ are free parameters.
As far as CP phases are concerned, we adopt, without loss of generality, the
convention ${\rm Arg}(\mu)=0$, while we assume a common phase for all
the $A_f$ terms, $\Phi_A\equiv {\rm Arg}(A_t)={\rm Arg}(A_b)={\rm
Arg}(A_\tau)$.  As a consequence of this, we end--up with two free
physical phases: $\Phi_A$ and $\Phi_3={\rm Arg}(M_3)$.

In addition to the parameters fixed by the CPX scenario,
we need to fix the gaugino masses $M_{1,2}$ for our study. 
We take them as free parameters
independently of $M_3$ since, for them, we chose to relax the usual
relations at the electro-weak scale:
$M_i/M_j=g_i^2/g_j^2$ with $g_{i,j}$=gauge coupling constants, 
which originate from the
assumption of gaugino--mass unification at the GUT scale.  
The neutralino $\chi$ is defined as usual as the lowest-mass linear
superposition of  $B$-ino $\tilde{B}$, $W$-ino $\tilde{W}^{(3)}$, and of the
two Higgsino states $\tilde{H}^0_1$, $\tilde{H}^0_2$:

\begin{equation}
\chi\equiv a_1 \tilde{B}+a_2\tilde{W}^{(3)}+a_3 \tilde{H}^0_1+a_4 \tilde{H}^0_2.
\end{equation}

\noindent

In Ref.~\cite{light_neutralinos} it was proved that in a
CP--conserving effective MSSM with $|M_1| << |M_2|$ light neutralinos
of a mass as low as 7 GeV are allowed. Indeed, for $|M_1| << |M_2|$
the LEP constraints do not apply, and the lower bound on the
neutralino mass is set by the cosmological bound. 
 In the following we will assume vanishing phases for
$M_1$ and $M_2$, and we will fix for definiteness $M_2$=200 GeV (the
phenomenology we are interested in is not sensitive to these
parameters in a significant way). On the other hand, we will vary
$M_1$, which is directly correlated to the lightest neutralino mass
$m_{\chi}$.

In this work, we rely on {\tt CPsuperH} ~\cite{CPsuperH} for the computation of
mass spectra and couplings in the MSSM Higgs sector.

\section{Experimental constraints on the CPX scenario}

In the CPX scenario the lightest Higgs boson is mostly CP odd and its
production at LEP is highly suppressed since $|g_{H_1 VV}|\ll 1$
though it is kinematically accessible. As a consequence of this,
taking $\Phi_A=\Phi_3=90^\circ$ and $M_{\rm SUSY}=0.5$ TeV, the
combined searches of the four LEP collaborations at $\sqrt{s}= 91 -
209$ GeV reported the following allowed interval for a very light
$H_1$~\cite{LEP_HIGGS}, which we will focus on in our analysis:
\begin{equation}
M_{H_1} \lsim 10 \;\; {\rm GeV}\;\;\;\; 
{\rm for}\;\; 3 \lsim \tan\beta \lsim 10.
\label{eq:lep}
\end{equation}
\noindent
We observe that in the scenario analyzed by the LEP collaborations one
has $|\mu|$=2 TeV. For this large value of $|\mu|$, the neutralino is
a very pure $B$-ino configuration, with a Higgsino contamination
$a_3\simeq$0.02. As will be shown in Section
\ref{section:relicdensity}, this has important consequences for the
phenomenology of relic neutralinos, in particular suppressing their
annihilation cross section, and restricting the possibility of having
a relic abundance in the allowed range only to the case of resonant
annihilation. So, the exploration of different possibilities with
lower values of $|\mu|$ could in principle be very relevant for relic
neutralinos.  However, this would require a re--analysis of LEP data
which is beyond the scope of this paper.

\begin{figure}[t]
\vspace{0.0cm}
\centerline{
\includegraphics[height=7cm,bb=23 47 490 488]{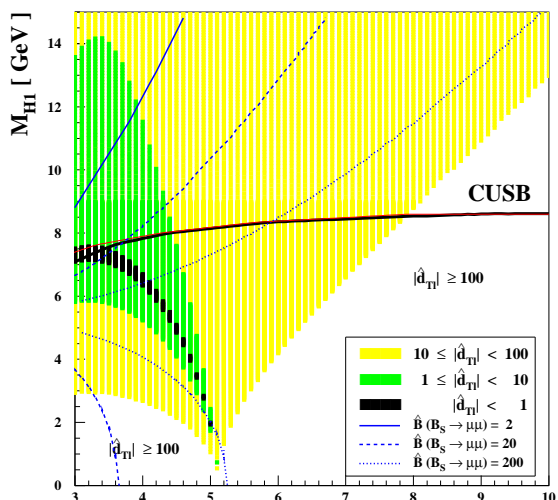}
}
\vspace{-0.5cm}
\caption{{\it The Thallium EDM $\hat{d}_{\rm Tl} \equiv d_{\rm Tl}
\times 10^{24}$~$e\,cm$ in the CPX scenario with $M_{\rm SUSY}=0.5$
TeV in the region $M_{H_1} \lsim 15$ GeV and $3 < \tan\beta < 10$.
The different shaded regions correspond to different ranges of
$|\hat{d}_{\rm Tl}|$, as shown: specifically, the narrow region
consistent with the current thallium EDM constraint, $|\hat{d}_{\rm
Tl}|<1$, is denoted by black squares.  In the blank unshaded region we
have $|\hat{d}_{\rm Tl}|>100$. The region below the thick solid line
is excluded by data on $\Upsilon(1S)$ decay. For comparison, the thin
line shows an estimation of the same boundary obtained using the
tree-level coupling taking $O_{a1}$=1, i.e. $|g^P_{H_1\bar{b}b}| =
\tan\beta$. Also shown are the three contour lines of the rescaled
$\hat{B}(B_s\rightarrow \mu\mu)\equiv (B_s\rightarrow \mu\mu) \times
10^7$: $\hat{B}(B_s\rightarrow \mu\mu)=2$ (solid), 20 (dotted), and
200 (dashed).}}
\label{fig:dtl}
\end{figure}

CP phases in the MSSM are significantly constrained by the EDM
measurements. In particular, the EDM of the Thallium atom provides
currently the most stringent constraint on the MSSM scenario of our
interest.  In Fig.~\ref{fig:dtl}, we show the rescaled Thallium EDM
$\hat{d}_{\rm Tl} \equiv d_{\rm Tl} \times 10^{24}$ in units of
$e\,cm$ in the $M_{H_1}$-$\tan\beta$ plane. Here, we consider only the
contributions from the Higgs-mediated two--loop diagrams\cite{KL_PR}.
Different ranges of $|\hat{d}_{\rm Tl}|$ are shown explicitly by
different shadings.  In the blank unshaded region we have obtained
$|\hat{d}_{\rm Tl}|>100$. We also note that the Thallium EDM
constraint can be evaded by assuming cancellations between the
two--loop contributions and other contributions, such as those from
first-- and second--generation sfermions.
In this way the allowed region shown in Fig.~\ref{fig:dtl} can be
enlarged.  The amount of cancellation can be directly read--off from
Fig.~\ref{fig:dtl}.  For instance, in the region $|\hat{d}_{\rm
Tl}|<10$ it would be less severe than 1 part in 10.

In the region of Eq.~(\ref{eq:lep}) the bottomonium decay
channel $\Upsilon(1S)\rightarrow \gamma H_1$ is kinematically
accessible ~\cite{higgs_hunter_guide}, and the experimental upper
bounds on this process can be directly converted to a constraint in
the plane $\tan\beta$--$M_{H_1}$.  The result is shown in
Fig.~\ref{fig:dtl}, where the thin (red) line corresponds to the limit
obtained by neglecting finite--threshold corrections induced by gluino
and chargino exchanges, and by setting $O_{a1}=1$, while the thick
solid line shows the same constraint when threshold corrections and
the true value of $O_{a1}$ are used.
From Fig.~\ref{fig:dtl} one can see that, when the following
constraints are combined: (i) the LEP constraint; (ii) Thallium EDM; 
(iii) the limit from bottomonium decay,
the allowed parameter space is reduced to:
\begin{equation}
7 \;\;{\rm GeV} \lsim  M_{H_1}\lsim 7.5 \;\;{\rm GeV}
~~{\rm and}~~ \tan\beta\simeq 3.
\label{eq:tanb_mh1}
\end{equation}
\noindent This region may be enlarged to
\begin{equation}
7 \;\;{\rm GeV} \lsim  M_{H_1}\lsim 10 \;\;{\rm GeV}
~~{\rm and}~~ 3\lsim \tan\beta\lsim 5\,,
\label{eq:tanb_mh1_enlarged}
\end{equation}
\noindent if we assume 10 \%-level cancellation in the
Thallium EDM.

In light of the above discussion and for definiteness, from now on we
will fix $M_{H_1}$=7.5 GeV and $\tan\beta$=3 in our analysis. Taking
into account the CPX parameter choice of Eq.(\ref{eq:mssm-intro-CPXdef})
with  $\Phi_A=\Phi_3=90^\circ$ and $M_{\rm SUSY}=0.5$ TeV,
this implies, in particular: $M_{H^{\pm}}\simeq$ 147 GeV,
$M_{H_2}\simeq$ 108 GeV, $M_{H_3}\simeq$ 157 GeV.

As will be discussed in the following sections, if the pseudoscalar
Higgs boson mass is in the range (\ref{eq:tanb_mh1}), a CPX light
neutralino with $m_{\chi}\lsim M_{H_1}/2$ can be a viable DM
candidate.  Due to their very pure $B$-ino composition, and to the
quite low value of $\tan\beta$, neutralinos in this mass range evade
constraints coming from accelerators. For instance, in the CPX light
neutralino mass range the present upper bound to the invisible width
of the $Z$--boson implies $|a_3^2-a_4^2|\lsim$ a few percent, a
constraint easily evaded in this case.

A potentially dangerous constraint can come from the decay
$B_s\rightarrow \mu\mu$, since its dominant SUSY contribution scales
as $\tan^6\beta\,|\mu|^2/M_{H_1}^4$ and may have a resonance
enhancement when $H_1$ is so light that $M_{H_1}\sim M_{B_s}$.
%
In Fig.~\ref{fig:dtl}, we show three contour lines of the rescaled
$\hat{B}(B_s\rightarrow \mu\mu)\equiv (B_s\rightarrow \mu\mu) \times
10^7$: $\hat{B}(B_s\rightarrow \mu\mu)=2$ (solid), 20 (dotted), and
200 (dashed).
For the parameters chosen by combining the results from LEP2 searches,
Thallium EDM, and Bottomonium decay, Eq. (\ref{eq:tanb_mh1}), we get:
$B(B_s\rightarrow \mu\mu)_{CPX}\simeq 6\times 10^{-7}$ taking
$f_{B_s}=0.23$ GeV.  This is three times larger than the present 95 \%
C.L. limit: $B(B_s\rightarrow \mu\mu)<2\times
10^{-7}$.  This can be easily made consistent with the present
experimental constraint if some mild cancellation takes place.  The ``GIM
operative point'' mechanism discussed in Ref.~\cite{dedes} may be an
example of such cancellation, when the squark mass matrices are
flavour diagonal. In particular, we find that $B(B_s\rightarrow
\mu\mu)_{CPX}$ is consistent to the experimental upper bound by
choosing $0.8\lsim \rho\lsim 0.9$, where $\rho\equiv
m_{\tilde{q}}/M_{\rm SUSY}$ is the hierarchy factor introduced in
Ref.~\cite{dedes}, with $m_{\tilde{q}}$ the soft mass for squarks of
the first two generations. 

For the discussion of other constraints, see \cite{scopel_lee}.

\section{The relic density}
\label{section:relicdensity}
Taking into account the latest data from the cosmic microwave data
(CMB) combined with other observations~\cite{wmap3} the 2--$\sigma$
interval for the DM density of the Universe (normalized to the
critical density) is:
\begin{equation}
0.096<\Omega_m h^2<0.122\,,
\label{eq:wmap3}
\end{equation}
\noindent where $h$ is the Hubble parameter expressed in units of 100
km s$^{-1}$ Mpc$^{-1}$. In Eq.(\ref{eq:wmap3}) the upper bound on
$\Omega_m h^2$ establishes a strict upper limit for the abundance of
any relic particle.  
In absence of some resonant effect, the natural scale of the
annihilation cross section times velocity $\sigma_{ann} v$ of CPX
light neutralinos is far too small to keep the relic abundance below
the upper bound of Eq.(\ref{eq:wmap3}) (in particular they are very
pure $B$--inos and their mass is below the threshold for annihilation
to bottom quarks, which is usually the dominant channel of
$\sigma_{ann} v$ for light neutralinos
~\cite{light_neutralinos}). However, when $m_{\chi}\simeq M_{H_1}/2$
neutralinos annihilate through the resonant channel
$\chi\chi\rightarrow H_1\rightarrow standard\; particles$, bringing
the relic abundance down to acceptable values. In the Boltzmann
approximation the thermal average of the resonant $\sigma_{ann} v$ to
the final state $f$ can be obtained in a straightforward way from the
following relation among interaction rates:
\begin{eqnarray}
&&\frac{n_{\chi}^2}{2}<\sigma_{ann} v>_{{\rm res},f} =
  <\Gamma(\chi\chi\rightarrow f)>=\nonumber\\
&&<\Gamma(\chi\chi\rightarrow
  H_1)B(H_1\rightarrow f)>=\nonumber\\
&&n_{H_1}\Gamma_{\chi}\frac{K_1(x_{H_1})}{K_2(x_{H_1})}B_f,
\label{eq:resonant}
\end{eqnarray}
\noindent where brackets indicate thermal average, $\Gamma_{\chi}$ is
the zero--temperature $H_1$ annihilation amplitude to neutralinos and
the thermal average of this quantity is accounted for by the ratio of
modified Bessel functions of the first kind $K_1$ and $K_2$, $B_f$ is
the $H_1$ branching ratio to final state $f$, $n_i=g_i m_i^3
K_2(x_i)/(2 \pi^2 x_i)$ are the equilibrium densities with
$x_i=m_i/T$, $T$ the temperature, and $g_i$ the corresponding internal
degrees of freedom, $g_{\chi}=2$, $g_{H_1}$=1. The factor of 1/2 in
front of Eq.(\ref{eq:resonant}) accounts for the identical initial
states in the annihilation.  

\begin{figure}[t]
\vspace{0.0cm}
\centerline{
\includegraphics[width=8cm,bb=17 175 504 666]{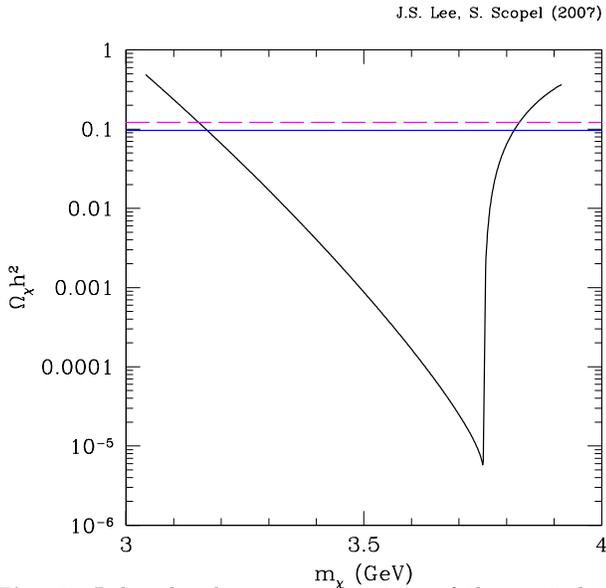}
}
\vspace{-0.5cm}
\caption{{\it Relic abundance as a function of the neutralino mass
$m_{\chi}$ for the CPX scenario with $M_{H_1}$=7.5 GeV, $\tan\beta$=3,
$M_{\rm SUSY}$=0.5 TeV and $\Phi_{A}=\Phi_3=90^\circ$.  The two
horizontal lines indicate the interval of Eq.(\protect\ref{eq:wmap3}).
\label{fig:mchi_omega}}}
\end{figure}

The result of our calculation is shown in Fig.~\ref{fig:mchi_omega},
where the neutralino relic abundance $\Omega_{\chi} h^2$ is shown as a
function of the mass $m_{\chi}$.  The asymmetric
shape of the curve in Fig.~\ref{fig:mchi_omega} is due to the fact that
thermal motion allows neutralinos with $m_{\chi}<M_{H_1}/2$ to reach
the center--of--mass energy needed to create the resonance, while this
is not possible for $m_{\chi}>M_{H_1}/2$. In the same figure, the two
horizontal lines indicate the range of Eq.(\ref{eq:wmap3}). 

In Fig.~\ref{fig:mchi_omega} the neutralino mass range allowed by
cosmology is: 3.15 GeV $\simeq m_{\chi} \simeq$ 3.83 GeV. Allowing for
the variation of $M_{H_1}$ within the range of
Eqs.(\ref{eq:tanb_mh1},\ref{eq:tanb_mh1_enlarged}), this range is
enlarged to:
\begin{equation}
2.93 \;\;{\rm GeV} \lsim m_{\chi}\lsim 5\;\; {\rm GeV}.
\label{eq:mass_range}
\end{equation}
\noindent In this scenario the neutralino relic abundance can fall in
the range of Eq.(\ref{eq:wmap3}) only with some level of tuning at the
boundaries of the allowed mass range. For intermediate values of
$m_{\chi}$ either the neutralino is a sub--dominant component of the
DM, or some non--thermal mechanism for its cosmological density needs
to be introduced.  Of course all our considerations are valid if
standard assumptions are made about the evolution of the early
Universe (e.g. about the reheating temperature at the end of
inflation, the energy budget driving Hubble expansion, entropy
production, etc).

\section{Dark matter searches}

\label{section:dmsearches}

Neutralinos in the halo of our Galaxy can be searched for through
direct and indirect methods. In particular, due to their very low mass
and cross section, CPX light neutralinos are quite hard to detect
through direct detection, although one proposal exists for such a hard
task \cite{collar}. As far as indirect searches of CPX light
neutralinos are concerned, in our scenario the neutralino relic
density $\Omega_{\chi} h^2$ is driven below the observational limit by
the resonant enhancement of the annihilation cross section
$\widetilde{<\sigma_{ann} v>}$.  The same cross section calculated at
present times, $<\sigma_{ann} v>_0$, enters into the calculation of
the annihilation rate of neutralinos in our galaxy. This could produce
observable signals, like $\gamma$'s, $\nu$'s or exotic components in
Cosmic Rays (CR), like antiprotons, positrons, antideuterons.  Note,
however, that one can have $<\sigma_{ann}
v>_0\ll\widetilde{<\sigma_{ann} v>}$. In fact, as already shown in
Section \ref{section:relicdensity}, the thermal motion in the early
Universe ($x_{\chi}\simeq x_f\simeq 20$) allows neutralino resonant
annihilation when $m_{\chi}<M_{H_1}/2$. However, for the same
neutralinos the contribution of the resonance to $<\sigma_{ann} v>_0$
can be negligible at present times, since their temperature in the
halo of our Galaxy is of order $x_{\chi,0}\simeq$10$^{-6}\ll
x_f$. This implies that the annihilation cross section can be large
enough in the early Universe in order to provide the correct relic
abundance, but not so large at present times as to drive indirect
signals beyond observational limits.  As a consequence of this, in
\cite{scopel_lee} we have discussed in detail signals for indirect
Dark Matter searches and shown that they are compatible with the
present experimental constraints, as long as $m_\chi\lsim
M_{H_1}/2$. On the other hand, part of the range $m_\chi\gsim
M_{H_1}/2$ allowed by cosmology is excluded by antiproton fluxes.
Moreover, in \cite{scopel_lee} prospects of detection in future DM
searches have been discussed, showing that CPX neutralinos might
indeed produce a detectable signal. Finally, as far as a very light
Higgs boson $H_1$ is concerned, we observe that the LHC might not be
able to detect it, and a Super $B$ factory could thus be needed for
its observation\cite{lozano}.

\end{document}